\documentclass[aps,prb,twocolumn,superscriptaddress]{revtex4-2}

\usepackage{amsmath}
\usepackage{amsthm}
\usepackage{amssymb}
\usepackage{amsfonts}
\usepackage{graphicx}
\usepackage{romannum}
\usepackage{float}
\usepackage{placeins}
\usepackage{color}
\usepackage{soul}
\usepackage{appendix}

\begin{document}

\title{Spontaneous spin splitting and tunable valley polarization in a two-dimensional fully compensated ferrimagnet}

\author{Hongyan Lv}
\email[Corresponding author: ]{hylv@hfut.edu.cn}
\affiliation{School of Physics, Hefei University of Technology, Hefei 230009, China}

\author{Yueli Li}

\affiliation{School of Physics, Hefei University of Technology, Hefei 230009, China}

\author{Yunfan Zhang}

\affiliation{School of Physics, Hefei University of Technology, Hefei 230009, China}

\author{Jiayu Dai}

\affiliation{School of Physics, Hefei University of Technology, Hefei 230009, China}

\author{Zhongjun Li}

\affiliation{School of Physics, Hefei University of Technology, Hefei 230009, China}

\author{Ding-Fu Shao}
\email[Corresponding author: ]{dfshao@issp.ac.cn}
\affiliation{Key Laboratory of Materials Physics, Institute of Solid State Physics, Chinese Academy of Sciences, Hefei 230031, China}

%%%%%%%%%%%%%%%%%%%%%%%%%%%%%% LyX specific LaTeX commands.

\begin{abstract}
Materials with controllable valley polarization and anomalous valley Hall (AVH) effect are highly desired in valleytronic applications. While current AVH studies primarily focus on ferromagnetic materials, two-dimensional (2D) antiferromagnets are more attractive for valleytronics since they possess zero net magnetization, negligible stray fields, and ultrafast spin dynamics. Nevertheless, the joint space-inversion and time-reversal ($PT$) symmetry in conventional collinear antiferromagnets prohibits the occurrence of AVH response. The recently proposed fully compensated ferrimagnets break $PT$ symmetry, and the spin-opposite sublattices are not related by crystal symmetry, providing a natural platform for the coexistence of spontaneous spin splitting, valley polarization, and anomalous-Hall compatible symmetry. Herein, we demonstrate that such compensated ferrimagnetism can be realized in a Janus Mn$_{2}$BrI monolayer, with a N\'{e}el temperature above room temperature. Spontaneous spin splitting is observed due to the built-in layer-dependent electrostatic potential. When SOC is considered, valley polarization emerges for an out-of-plane N\'{e}el vector. Moreover, proper hole doping stabilizes the perpendicular magnetic anisotropy and the two valleys exhibit markedly different Berry curvatures, thereby making AHE responses allowed. Furthermore, the valence band extrema of Mn$_{2}$BrI monolayer can be effectively tuned by external biaxial strain and giant piezomagnetism can be achieved. Our results identify Janus Mn$_{2}$BrI monolayer as a promising fully compensated ferrimagnetic platform for 2D valleytronics and spintronics.

\end{abstract}
\pacs{73.22.-f, 75.30.-m, 75.30.Et}

\maketitle
\pagenumbering{arabic}

\section{INTRODUCTION}

Valley refers to an energy extremum in the electronic band structure in momentum space, such as a local conduction-band minimum or valence-band maximum, and constitutes an electronic degree of freedom distinct from charge and spin. Because inequivalent valleys can in principle be selectively generated, manipulated, and detected, valleytronics has been widely regarded as a promising route for information storage and processing \cite{valley1,valley2,valley3,D.Xiao-PRL-2012,X.Xu-NatRevMat-2016}. For practical valleytronic applications, a material platform should ideally host well-defined valleys, support controllable valley polarization, and, for device implementation, provide a direct electrical readout of the valley state. In this respect, anomalous valley Hall (AVH) effects are particularly attractive because they can convert valley polarization into a measurable transverse voltage signal.

A prototypical route toward such functionality is offered by ferrovalley materials \cite{VSe2-ferrovalley}, in which spontaneous valley polarization arises from the coexistence of broken space inversion ($P$) symmetry, spin–orbit coupling (SOC), and magnetic order. Most ferrovalley candidates discussed so far are ferromagnets with a finite net magnetization \cite{P.Zhao-APL-2019,X.Y.Feng-PRB-2021,R.Peng-PRB-2020}. While this makes valley control and electrical readout conceptually straightforward, ferromagnetic materials also suffer from intrinsic drawbacks, including stray fields, sensitivity to magnetic perturbations, and relatively slow spin dynamics. 

By contrast, antiferromagnets are attractive for valleytronics because they possess vanishing net magnetization, negligible stray fields, and ultrafast spin dynamics \cite{Spintronics-RevModPhys-2004,AFM-NatPhys-2018}. In conventional collinear antiferromagnets with broken space-inversion ($P$) and time-reversal ($T$) symmetries but preserved $PT$ symmetry, valley polarization may arise in suitable crystal structures because the two valleys are no longer related by either $P$ or $T$ \cite{PNAS110-2013-Li}. However, $PT$ symmetry imposes strong constraints on Berry curvature and transverse charge transport, such that a net AVH response is forbidden. Several methods have been proposed to break $PT$ symmetry in conventional antiferromagnets, such as constructing heterostructures to introduce a spatially nonuniform potential, modulating stacking configurations, and applying an out-of-plane electric field \cite{npj-Ma-2022,MaterHori-Ma-2023,APL-Meng-2025,PRB-Guo-2024-1,PRB-Guo-2024-2}. 

Recently, altermagnets have emerged as a special class of antiferromagnets that exhibit momentum-dependent spin splitting even in the absence of SOC, giving rise to nonrelativistic spin-transport characteristics \cite{PRX12-2022-1,PRX12-2022-2,Yao-AFM,NatCommun-V2Se2O-2021,NanoLett-V2SeTeO-2024,AdvMater-Cr2Se2O-2024,NanoLett-Fe2Se2O-2024,APL124-2024,PRL-Ca(CoN)2-2024,JPCM-M2X2O-2025}. However, although altermagnets do not preserve $PT$ symmetry, their spin-opposite sublattices are still connected by crystallographic symmetry operations, and altermagnetism by itself does not guarantee either valley polarization or AVH transport. A more flexible platform may instead be provided by fully compensated ferrimagnets \cite{PRL74-1995,PRB93-2016,PRL132-2024,PRL134-2025}, in which the two spin-opposite sublattices are not related by symmetry operations. This absence of symmetry linkage provides a natural platform for the coexistence of spontaneous spin splitting, valley polarization, and anomalous-Hall-compatible symmetry, making fully compensated ferrimagnets appealing for valleytronics.

Here, we demonstrate that such compensated ferrimagnetism can be realized in a Janus Mn$_{2}$BrI monolayer. By janusizing the parent A-type antiferromagnetic (AFM) monolayer, a built-in layer-dependent electrostatic potential is introduced, which breaks the symmetry protecting spin degeneracy and converts the system into a fully compensated ferrimagnet. We find that pronounced spontaneous spin splitting appears even in the absence of SOC. When SOC is included, valley polarization emerges for an out-of-plane N\'{e}el vector. Furthermore, proper hole doping stabilizes perpendicular magnetic anisotropy, and the two valleys exhibit markedly different Berry curvatures, thereby making anomalous valley Hall responses allowed. In addition, biaxial strain effectively tunes the valence-band extrema and gives rise to a giant piezomagnetic response. These results identify Janus Mn$_{2}$BrI monolayer as a promising fully compensated ferrimagnetic platform for two-dimensional valleytronics and spintronics.

\section{COMPUTATIONAL DETAILS}

The electronic and magnetic properties were investigated based on the density functional theory (DFT) as implemented in the Vienna $ab$ $initio$ simulations package (VASP) \cite{vasp1,vasp2}. The Perdew-Burke-Ernzerhof (PBE) expression \cite{PBE} of the generalized gradient approximation (GGA) was employed to describe the exchange-correlation functional. The core electrons were treated using the projector augmented wave method \cite{PAW}. The plane-wave cutoff energy was set to be 700 eV. A vaccum space larger than 25 {\AA} was added along the $c$ axis to avoid the interactions between each monolayer and its period image. The first Brillouin zone was sampled using a $\Gamma$ centered $18\times18\times1$ Monkhorst-Pack $k$-point mesh. Structural relaxation was stopped until the force acting on each atom was less then 0.01 eV/{\AA}. To treat the strong correlations of Mn 3$d$ electrons, DFT plus on-site Coulomb interaction $U$ (DFT+$U$) with the Dudarev approach was used \cite{Dudarev}.

The phonon spectrum was calculated based on the density functional perturbation theory (DFPT). A $6\times6\times1$ supercell and a $\Gamma$ centered $4\times4\times1$ Monkhorst-Pack $k$-point mesh were used. The phonon frequencies were obtained by the PHONOPY code \cite{phonopy}. The thermal stability was evaluated by performing $ab$ $initio$ molecular dynamics (AIMD) simulations \cite{AIMD}, using the Nos\'{e}-Hoover thermostat for 8 ps with a time step of 3 fs at 300 K in a $4\times4\times1$ supercell.

The magnetic exchange interactions were calculated by first formulating a tight-binding Hamitonian in the basis of maximally localized Wannier functions (MLWFs) \cite{Wannier1,Wannier2} and then employing the TB2J package \cite{TB2J}, which is based on the Green's functions. The spin Hamiltonian has the following form:
\[\displaystyle H = - \sum_{i \textless j}{J_{ij}\vec{S}_{i}\cdot\vec{S}_{j}} - \sum_{i}{A_i{S}_{iz}^{2}} - \sum_{i \textless j}{\vec{D}_{ij}\cdot(\vec{S}_i\times\vec{S}_j)},\]
where $J_{ij}$ represents the isotropic exchange interaction between magnetic moments at sites $i$ and $j$, and $\vec{S}_{i}$ and $\vec{S}_{j}$ are corresponding unit vectors denoting the directions of local magnetic moments. $A_{i}$ accounts for single-ion anisotropic energy and $\vec{D}_{ij}$ represents the Dzyaloshinshii-Moriya (DM) interaction. The magnetic transition temperatures were then calculated by solving the atomistic Landau-Lifshitz-Gilbert (LLG) equation, as implemented in the VAMPIRE package \cite{VAMPIRE}. The Mn$_{2}$BrI monolayer with a size of $60\times60$ nm was used in the simulation, with 2 000 000 equilibration and averaging steps using a time step of $5\times10^{-4}$ fs.

\section{RESULTS AND DISCUSSION}

\subsection{Spin splitting and valley polarization}

\begin{figure}
\includegraphics[width=1.0\columnwidth]{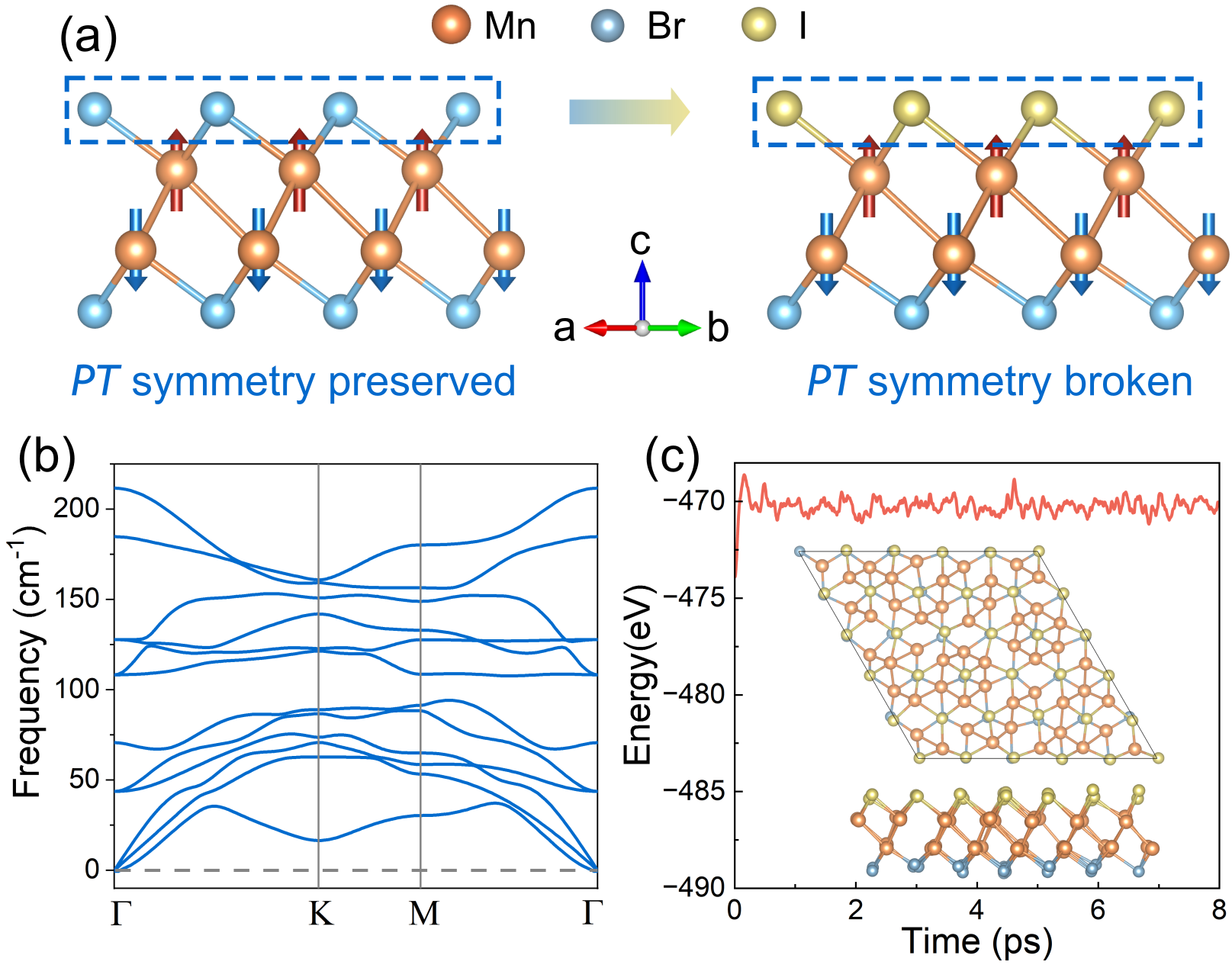}\caption{\label{Fig1} (a) Side views of Mn$_{2}$Br$_{2}$ (left panel) and Mn$_{2}$BrI (right panel) monolayers. (b) Phonon spectrum of Mn$_{2}$BrI monolayer. (c) Evolution of total energy as a function of simulation time for Mn$_{2}$BrI monolayer. Insets are top and side views of snapshot of Mn$_{2}$BrI monolayer from AIMD simulation at 8 ps.}
\end{figure}

\begin{figure*}
	\includegraphics[width=1.7\columnwidth]{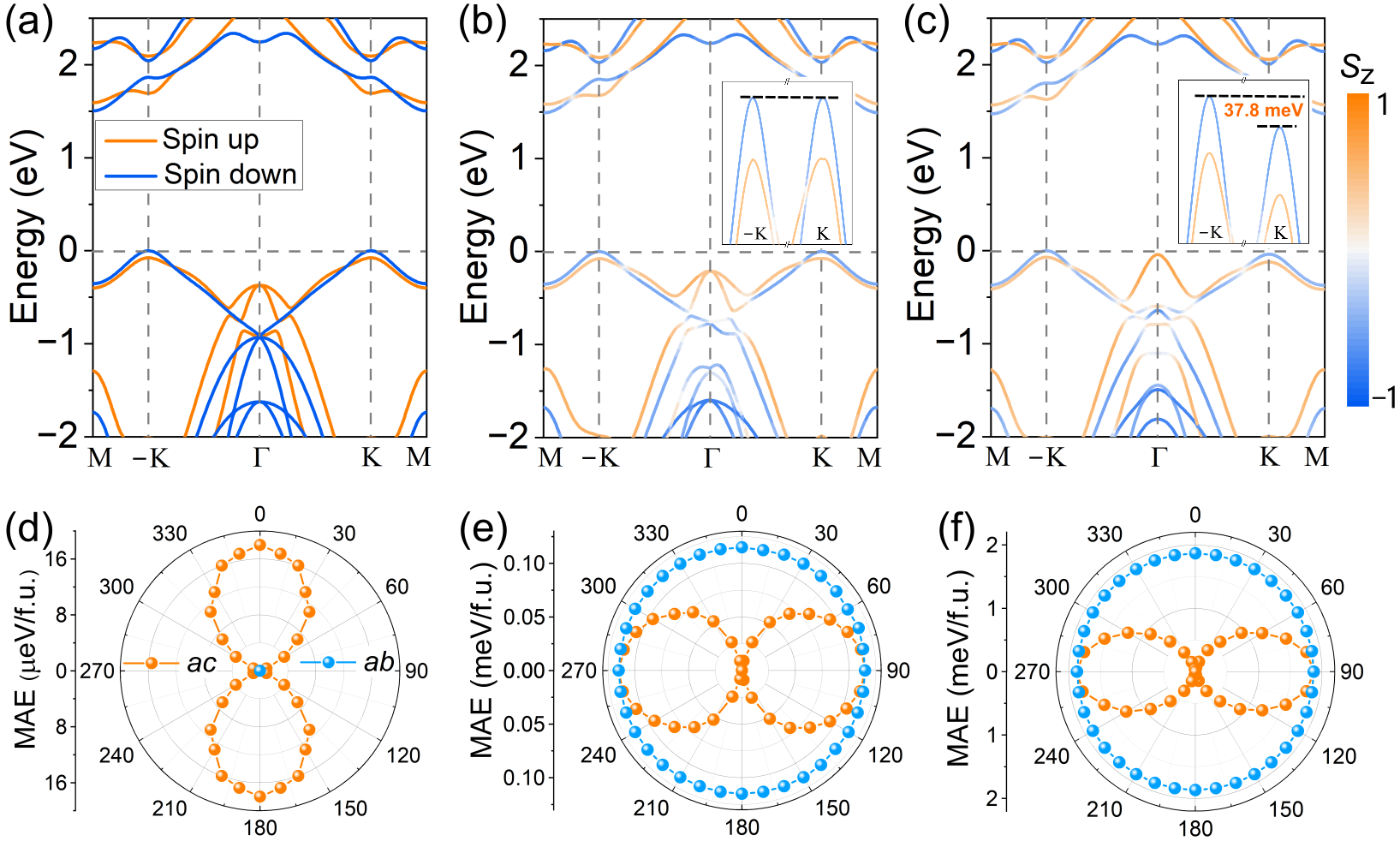}\caption{\label{Fig2} Band structures of Mn$_{2}$BrI monolayer (a) without spin-orbital coupling (SOC) and with SOC considered for N\'{e}el vector along (b) $a$ and (c) $c$ directions. Insets in (b) and (c) are the enlarged conduction bands in the vicinity of Fermi energy near $-$K and K points. Magnetic anisotropy energy (MAE) for (d) pristine, (e) 0.02 hole/f.u. doped and (f) 0.02 hole/f.u. doped with $-2\%$ strained Mn$_{2}$BrI monolayers.}
\end{figure*}

The Mn$_{2}$X$_{2}$ (X=Br and I) monolayer is composed of two hexagonal Mn layers sandwiched by two hexagonal halogen layers in the stacking order of X-Mn-Mn-X [left panel of Fig. 1(a)], which belongs to the family of transition metal chloride \cite{MX-ACR-1981}. The Janus Mn$_{2}$BrI monolayer can be constructed by replacing one of the two Br layers in Mn$_{2}$Br$_{2}$ monolayer by I atoms [right panel of Fig. 1(a)]. The non-Janus Mn$_{2}$Br$_{2}$ and Mn$_{2}$I$_{2}$ monolayers have the space group of $P$\={3}$m$1 (no. 164), possessing space inversion ($P$) symmetry. While for Janus Mn$_{2}$BrI monolayer, the $P$ symmetry is broken and the space group becomes $P$3$m$1 (no. 156). The dynamical stability of Mn$_{2}$BrI monolayer was confirmed by calculating the phonon spectrum, which shows no imaginary frequencies [Fig. 1(b)]. We further verify the thermal stability of Mn$_{2}$BrI monolayer by AIMD simulation at 300 K, which shows minimal alterations in structure and only slight energy fluctuations up to 8 ps [Fig. 1(c)].

To determine the magnetic ground state of Mn$_{2}$BrI monolayer, FM and three different AFM spin configurations are considered, as shown in Fig. S1 in the Supplemental Material \cite{SM}. The relative total energies of the different spin configurations are shown in Fig. S2. The influence of effective Hubbard $U$ value ($U_{\scriptsize\mbox{eff}}$) on the energy difference was also considered, which shows that the AFM1 configuration always has the lowest total energy with the values of $U_{\scriptsize\mbox{eff}}$ ranging from 0 to 5 eV (Fig. S2). The magnetic ground state of Mn$_{2}$BrI monolayer is AFM1, in which the intralayer Mn atoms are ferromagnetically coupled, while the interlayer ones are coupled antiferromagnetically. The band structures calculated using different values of $U_{\scriptsize\mbox{eff}}$ as well as employing the screened hybrid HSE approximation (HSE06) method \cite{HSE06} are shown in Fig. S3. The band structure calculated with a modest $U_{\scriptsize\mbox{eff}}$ of 4 eV is very close to that calculated using HSE06 functional, therefore, in the following, we choose to use PBE+$U$ method with $U_{\scriptsize\mbox{eff}}$ of 4 eV.

When considering the AFM1 spin configuration, both the space inversion ($P$) and time reversal ($T$) symmetries are broken for not only the Janus Mn$_{2}$BrI monolayer, but also the non-Janus counterparts of Mn$_{2}$Br$_{2}$ and Mn$_{2}$I$_{2}$ monolayers.  However, for the non-Janus systems, the combined $PT$ symmetry preserves, which ensures the Kramers degeneracy in these two systems and their band structures are spin degenerate in the absence of SOC (Fig. S4). Therefore, the magnetic ground states for the non-Janus systems are both A-AFM. While for Janus Mn$_{2}$BrI monolayer, the built-in electric field arising from the Janus structure breaks the $PT$ symmetry and spontaneous spin splitting is observed [Fig. 2(a)]. Consequently, by Janusization, the system changes to be a cFiM material. In the vicinity of Fermi level, the spin splitting can reach as high as 0.54 eV at $\Gamma$ point, larger than those induced by relativistic SOC effect \cite{SOC-induced-splitting}. The Mn$_{2}$BrI monolayer is an indirect band-gap semiconductor with a band gap of 1.5 eV. The magnetic anisotropy energy (MAE) is calculated with SOC included, which is determined by rotating the spins from $c$-axis in the $ac$-plane and from $a$-axis in the $ab$-plane, and the angle-dependent relative energies are plotted in Fig. 2(d). We can see that the pristine Mn$_{2}$BrI monolayer has an easy magnetisation plane ($ab$-plane). The energy with N\'{e}el vector along out-of-plane direction ($E_{001}$) is 18 $\mu$eV/f.u. higher than that when N\'{e}el vector is along the in-plane $a$-axis ($E_{100}$).

\begin{figure}
	\includegraphics[width=1.0\columnwidth]{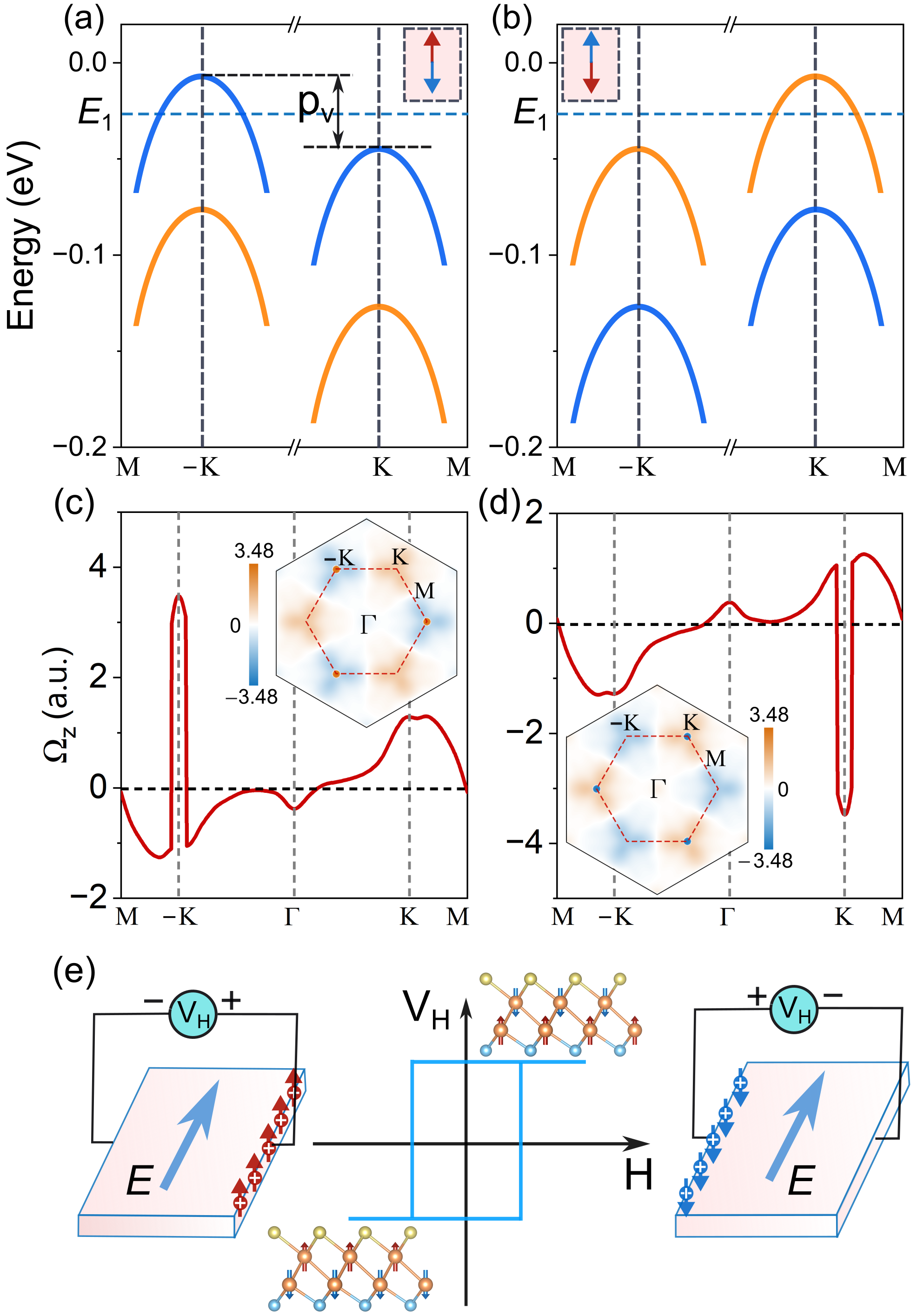}\caption{\label{Fig3} Schematic illustration of the valley polarization and spin splitting of top valence bands at $-$K and K points for Mn$_{2}$BrI monolayer at (a) m$\uparrow\downarrow$ and (b) m$\downarrow\uparrow$ states. Out-of-plane Berry curvatures of Mn$_{2}$BrI monolayer along high-symmetry paths when the Fermi energies ($E_{F}$) are moved to the middle of $-$K and K valleys [position of $E_{1}$ as indicated in (a) and (b)] for (c) m$\uparrow\downarrow$ and (d) m$\downarrow\uparrow$ states. Insets in (c) and (d) are the Berry curvature distributions across the Brillouin zone. (e) Schematic illustration of anomalous valley Hall (AVH) effect for Mn$_{2}$BrI monolayer in m$\uparrow\downarrow$ (left panel) and m$\downarrow\uparrow$ (right panel) states, and realization of magnetically writing and electrically reading memory device based on AVH effect (middle panel).}
\end{figure}

Next the SOC effect on the band structure is investigated. The band structures with N\'{e}el vector along $a$ and $c$ axes are shown in Figs. 2(b) and (c), respectively. We can see that a change of N\'{e}el vector orientation significantly modifies the band structure, exhibiting a magneto band-structure effect, similar to CrI$_{3}$ monolayer \cite{CrI3-Jiang}. To be specific, when the N\'{e}el vector is along the in-plane $a$-axis [Fig. 2(b)], the two top spin-up valence bands at the $\Gamma$ point in the vicinity of Fermi energy are slightly elevated, compared with the bands without SOC. When the N\'{e}el vector is along the out-of-plane $c$-axis [Fig. 2(c)], the above mentioned two spin-up bands are significantly split, one of which is elevated close to the Fermi level and the other moves downward. More importantly, when the spins are oriented along the out-of-plane direction, not only the spin splitting is maintained, but also a valley polarization emerges, that is, the energy degeneracy at $-$K and K valleys is lifted. The value of valley polarization ($P_{v}$) is defined as the energy difference between band extrema at $-$K and K valleys, i.e., $P_{v}=E$($-$K)$-E$(K), which is calculated to be 37.8 meV. Therefore, the Janus Mn$_{2}$BrI monolayer is a ferrovalley material when the N\'{e}el vector is oriented out-of-plane. 

Interestingly, we find that upon hole doping, the easy magnetization axis of Mn$_{2}$BrI monolayer will concurrently change to be out-of-plane [Figs. 2(e) and 4(a)] and the MAE increases monotonically as a function of the hole concentration. For hole doping of 0.05 hole/f.u., the MAE increases to be 0.21 meV/f.u., much larger than that of the pristine system. Therefore, for the hole-doped Mn$_{2}$BrI monolayer, the perpendicular MAE accompanied by valley polarization, as well as spin splitting coexist, which will bring in the possibility of AVH effect. For this case, the magnetic space group is $P$3$m^{\prime}$1 for Mn$_{2}$BrI monolayer, which ensures the occurrence of anomalous Hall conductivity (AHC) based on the Onsager relations \cite{Symmetr-2026,Stokes-JAC-2005,Gallego-JAC1-2016,Gallego-JAC2-2016}.

\begin{figure*}
	\includegraphics[width=1.7\columnwidth]{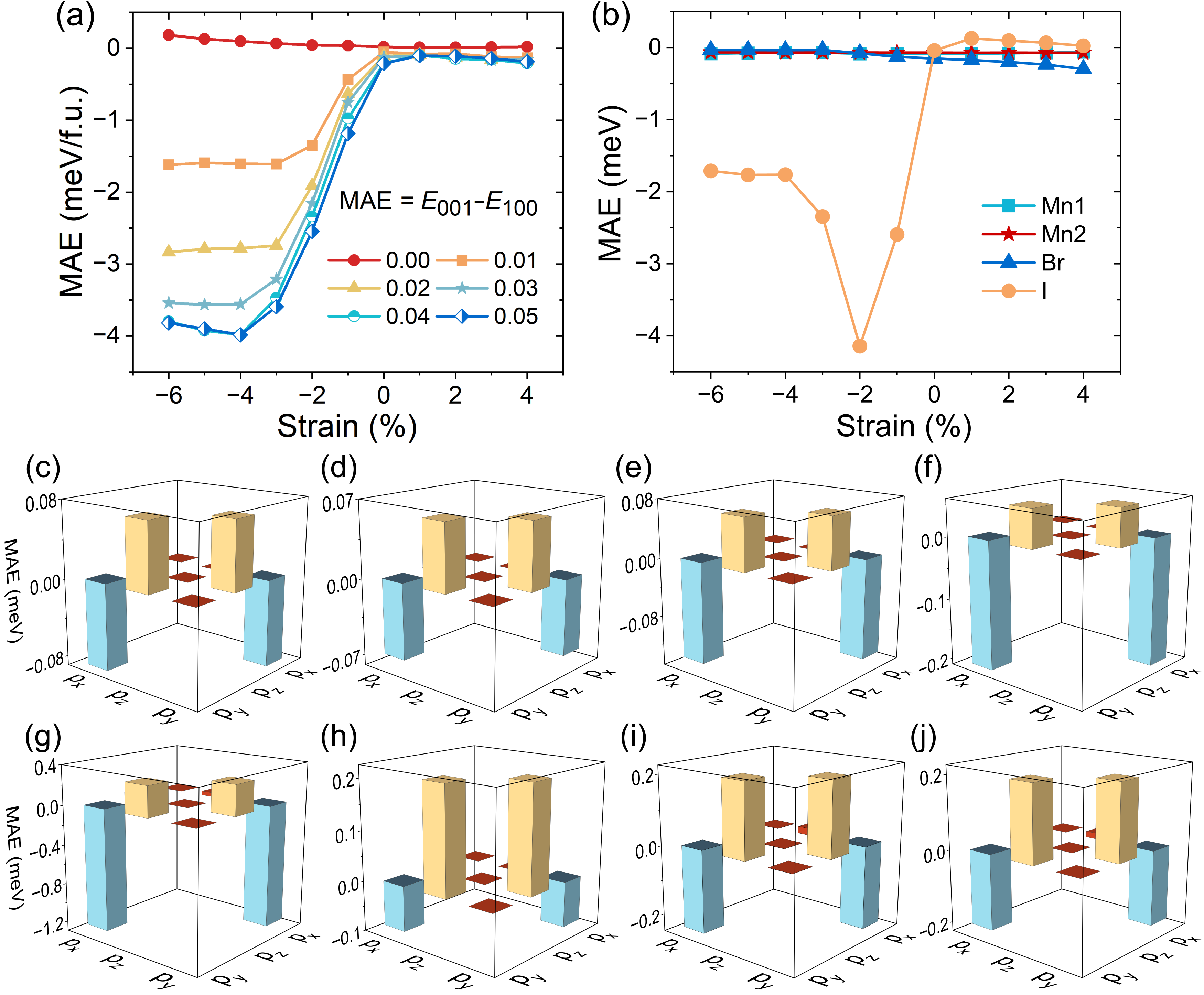}\caption{\label{Fig4} (a) Strain-dependent MAE (= $E_{001}-E_{100}$) for different hole concentrations, where $E_{001}$ and $E_{100}$ are the total energies when N\'{e}el vectors are along [001] and [100] directions, respectively. (b) Atom-resolved MAE for 0.02 hole/f.u. doped Mn$_{2}$BrI monolayer as a function of strain. The Br-3$p$-orbital-resolved MAE for (c) $-4$$\%$ strained and 0.02 hole/f.u. doped, (d) pristine, (e) 0.02 hole/f.u. doped, and (f) $4$$\%$ strained and 0.02 hole/f.u. doped Mn$_{2}$BrI monolayer. The I-3$p$-orbital-resolved MAE for (g) $-4$$\%$ strained and 0.02 hole/f.u. doped, (h) pristine, (i) 0.02 hole/f.u. doped, and (j) $4$$\%$ strained and 0.02 hole/f.u. doped Mn$_{2}$BrI monolayer.}
\end{figure*}

The intrinsic AHC is often related to the Berry curvature of the occupied bands \cite{Berry1,Berry2}. In the following, we further evaluate the Berry curvature of Mn$_{2}$BrI monolayer, which is defined as
\[\Omega(\mathbf{k})=-\sum_{n}\sum_{n\neq n^\prime}f_n\frac{2\mbox{Im}\langle\psi_{n\mathbf{k}}|v_x|\psi_{n^\prime\mathbf{k}}\rangle\langle\psi_{n^\prime\mathbf{k}}|v_y|\psi_{n\mathbf{k}}\rangle}{(E_n-E_{n^\prime})^2},
\]
where $f_{n}$ denotes the Fermi-Dirac distribution function, $\psi_{n\mathbf{k}}$ is the Bloch wave function with eigenvalue $E_{n}$, and $v_{x}$/$v_{y}$ is the velocity operator along $x/y$ direction. In 2D hexagonal lattices with inversion symmetry breaking, the $-$K and K valleys will exhibit a nonzero Berry curvature along the out-of-plane direction. For nonmagnetic transition metal dichalcogenides, MoS$_{2}$ for example, the Berry curvature has an odd parity $\Omega_{z}(\mathbf{k})$=$-\Omega_{z}(\mathbf{-k})$, due to the time-reversal symmetry. For the Mn$_{2}$BrI monolayer, the perpendicular MAE and valley polarization coexist when the system is hole doped. We calculate the out-of-plane Berry curvature $\Omega_{z}(\mathbf{k})$ when Fermi energy ($E_{F}$) is moved in the middle of $-$K and K valleys [position of $E_{1}$ as indicated in Fig. 3(a)], which is shown in Fig. 3(c). We can see that the signs of $\Omega_{z}$ near $-$K and K points are opposite in general, expect that there exists a maximum value with opposite sign at $-$K point, and thus the two valleys exhibit markedly different Berry curvatures. For this case, the upper Mn layer is aligned spin-up and the lower Mn layer is spin-down, corresponding to the m$\uparrow\downarrow$ state [Fig. 3(a)], and near $-$K and K valleys, the spin-down bands have larger energy than the spin-up bands. If the N\'{e}el vector of Mn$_{2}$BrI monolayer is reversed [m$\downarrow\uparrow$ state, Fig. 3(b)], the valley at K point has larger energy than that at $-$K point, that's to say, the valley polarization is reversed as well. In addition, the spin-up bands have higher energy than the spin-down bands, and the spin splitting is also reversed. In this case, the K valley has much larger Berry currature $\Omega_{z}$ than the $-$K valley if $E_{F}$ is moved between $-$K and K valleys [Fig. 3(d)].

If the Mn$_{2}$BrI monolayer is hole doped with a moderate concentration, shifting the Fermi energy between $-$K and K valleys in the valence band, the majority carriers acquire an anomalous transverse velocity $\vec{v}$ towards boundary of the sample in the presence of a longitudinal in-plane electric field $\vec{E}$, based on $\vec{v}\sim\vec{E}\times\Omega(\mathbf{k})$. When N\'{e}el vector points upward [Fig. 3(a)], the net spin-up holes from $-$K valley gain transverse velocities towards right side, generating a charge Hall current that can be detected as a negative voltage [left panel of Fig. 3(e)]. When the N\'{e}el vector turns downward, the reversal of valley polarization and spin spitting leads to accumulation of net spin-down holes at the left boundary, resulting in a positive voltage [right panel of Fig. 3(e)]. For system with both spin splitting and valley polarization, the net charge comes from the same valley with the same spin, realizing a simultaneous charge, spin, and valley polarization, termed as AVH effect. Based on the AVH effect, the magnetically writing and electrically reading memory devices can be realized [middle panel of Fig. 3(e)], and the reversal of N\'{e}el vector can be electrically detected in Mn$_{2}$BrI monolayer with fully compensated ferrimagnetism.

\subsection{Strain effect in combination of hole doping}

\subsubsection{Magnetic anisotropy energy}

Firstly, we will show that the MAE of Mn$_{2}$BrI monolayer can be further increased by hole-doping combined with the method of strain. Here, only the biaxial strain is considered, which is defined as $\epsilon$=$(a-a_{0})/a_{0}$, where $a_{0}$ and $a$ are the in-plane lattice constants of the Mn$_{2}$BrI monolayer before and after the strain is applied, respectively. The negative (positive) value corresponds to compressive (tensile) strain. As shown in Fig. 4(a), for the undoped Mn$_{2}$BrI monolayer, both the compressive and tensile strains keep the easy magnetization axis in-plane. For the compressive strain, the value of MAE increases as a function of the applied strain, while it is slightly decreased for the tensile strain.

\begin{figure*}
	\includegraphics[width=1.7\columnwidth]{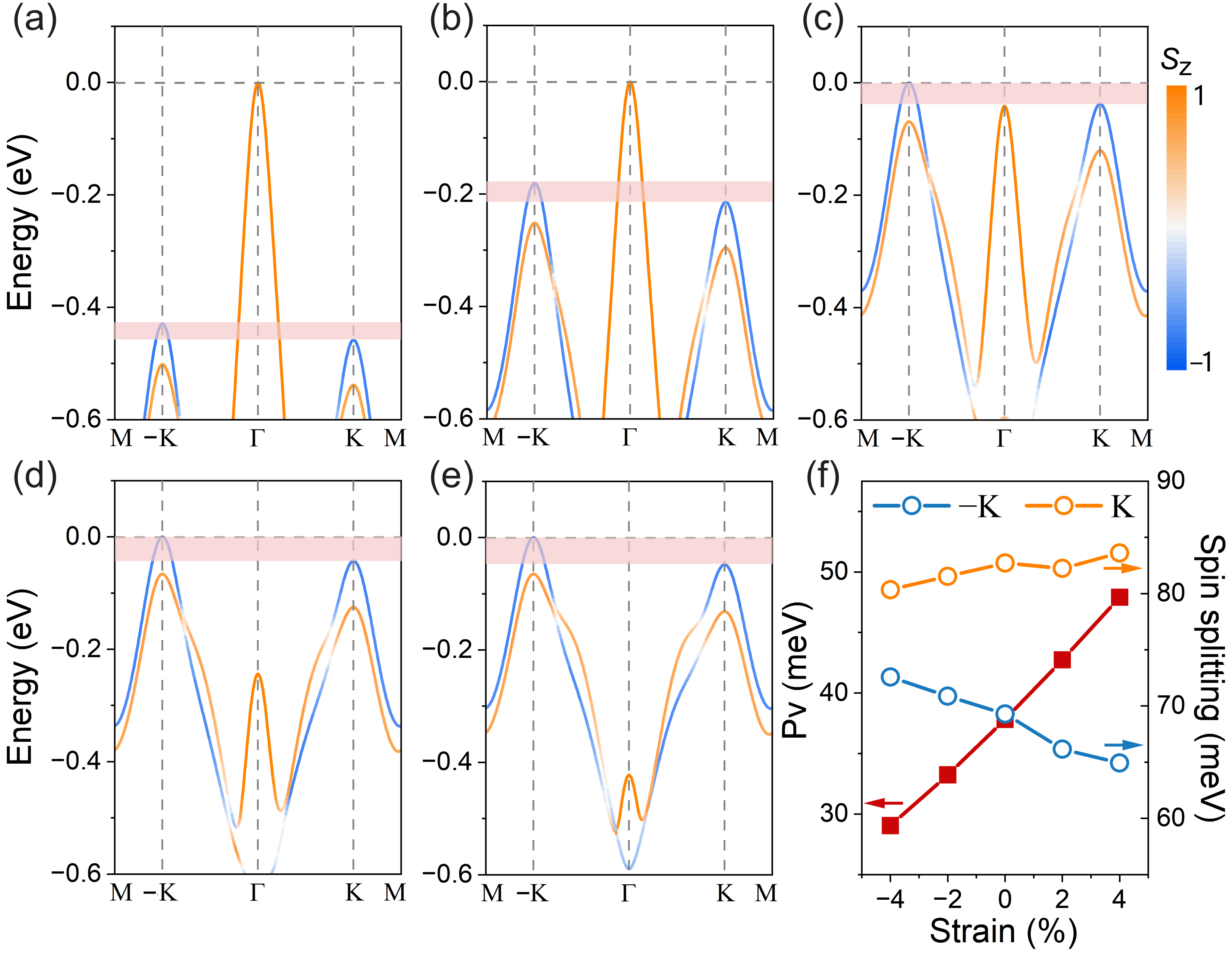}\caption{\label{Fig5} Band structures of Mn$_{2}$BrI monolayer when strains of (a) $-4$$\%$, (b) $-2$$\%$, (c) 0, (d) $2$$\%$, and (e) $4$$\%$ are applied, with SOC considered. (f) Valley polarization $P_{v}$ (red squares) and spin splitting at $-$K (blue circles) and K (orange circles) valleys as a function of strain.}
\end{figure*}

However, as long as the system is hole-doped, the easy magnetization axis will be tuned out-of-plane, not only for the unstrained Mn$_{2}$BrI monolayer, but also for the system with biaxial strain applied [Fig. 4(a)]. On one hand, for a particular hole concentration, we can see that MAE can be increased significantly by the applied compressive strain, while the tensile strain has little impact. For the relatively low concentrations of 0.01 and 0.02 hole/f.u., the absolute value of MAE ($|\mbox{MAE}|$) increases rapidly when the compressive strain is increased up to $-3$$\%$. When the strain is further increased, the $|\mbox{MAE}|$ will be slightly increased. For higher hole concentrations (0.03, 0.04, and 0.05 hole/f.u.), the overall trend is similar, except that the critical strain increases to $-4$$\%$. When the strain is further increased, the $|\mbox{MAE}|$ remains unchanged for hole doping of 0.03 hole/f.u. and are slightly decreased for hole concentrations of 0.04 and 0.05 hole/f.u. On the other hand, for a particular strain, the absolute value of MAE ($E_{001}-E_{100}$) increases monotonically as a function of hole concentration, the trend of which is the same as the unstrained case. For the strain and hole range we considered, the absolute value of MAE can reach as high as 3.98 meV/f.u. for system with compressive strain of $-4$$\%$ and hole doping of 0.05 hole/f.u.

To investigate the origin of the significantly increased MAE, we derive the atom-resolved MAE of Mn$_{2}$BrI monolayer. We take the system with hole doping concentration of 0.02 as a prototype and the results are shown in Fig. 4(b). We can see that for the compressive strain, the change of MAE is dominated by I atom, while for the tensile strain, both the I and Br atoms contribute to the MAE change. We then analyze the orbital-resolved MAEs of Br and I atoms and find that the $p$-orbitals play the main role in determining the change of MAE. In Fig. 4(c-j), we plot the p-orbital-resolved MAEs of Br and I atoms for $-4$$\%$ strained with 0.02 hole/f.u. doped, pristine, 0.02 hole/f.u. doped, and 4$\%$ strained with 0.02 hole/f.u. doped  Mn$_{2}$BrI monolayer. For pristine Mn$_{2}$BrI monolayer [Figs. 4(d) and (h)], the hybridization between $p_{x}$ and $p_{y}$ gives rise to perpendicular magnetic anisotropy (PMA), while the hybridization between $p_{y}$ and $p_{z}$ gives rise to in-plane magnetic anisotropy (IMA). For Br atom, the contributions from the PMA and IMA are comparable. While for I atom, the larger hybridization between $p_{y}$ and $p_{z}$ orbitals determines the IMA of pristine Mn$_{2}$BrI monolayer. When the system is hole-doped with concentration of 0.02 [Figs. 4(e) and (i)], for both Br and I atoms, the PMA contribution comes from the $p_{x}$ and $p_{y}$ hybridization is enhanced, exceeding the IMA contribution arising from the $p_{y}$ and $p_{z}$ hybridization. Therefore, the easy magnetization axis tends to be out-of-plane. However, for the unstrained system, although the easy magnetization axis can be tuned by hole doping, the increasement of $|\mbox{MAE}|$ is relatively small. For hole concentration of 0.05, the $|\mbox{MAE}|$ is enhanced to be 0.21 meV/f.u. 

For the compressive strain, however, the $|\mbox{MAE}|$ is significantly increased. Taking the 0.02 hole/f.u. doped system with strain of $-4\%$ for example [Figs. 4(c) and (g)], the PMA contribution coming from the $p_{x}$ and $p_{y}$ hybridization of I atom [Figs. 4(g)] dominates in determining the PMA property of Mn$_{2}$BrI monolayer. Therefore, the MAE can be significantly increased for Mn$_{2}$BrI monolayer with compressive strain in combination with hole doping. For the tensile strain, the Br atom dominates in determining the MAE. The $p_{x}$ and $p_{y}$ hybridization is larger than the $p_{y}$ and $p_{z}$ hybridization, but the difference is not as large as that for the compressive strain. Therefore, the increasement of MAE for hole doped system with tensile strain is relatively small.

\subsubsection{Valley polarization and anomalous valley Hall effect}

We further investigate the strain effect on the valley polarization of Mn$_{2}$BrI monolayer, as shown in Fig. 5. Since the easy magnetization axis tends to be out-of-plane upon hole doping for all the strain range considered, we plot the band structures when N\'{e}el vector is along out-of-plane direction. When the biaxial compressive strain is applied, the band extremum at $\Gamma$ point is elevated gradually as the strain is increased, while the band extrema at $-$K and K points shift downward [Figs. 5(a) and (b)]. The spin splitting at $-$K point is increased, while that at K point is decreased. The valley polarization $P_{v}$ is reduced for the compressive strain [Figs. 5(f)]. For the tensile strain, however, the band extremum at $\Gamma$ point moves downward, far away from $E_{F}$, while the valleys at $-$K and K points are elevated up to $E_{F}$ [Figs. 5(d) and (e)]. In addition, the valley polarization $P_{v}$ increases linearly as a function of strain, although the spin splitting at $-$K point is decreased. At the strain of 4$\%$, the valley polarization reaches 47.91 meV, and the spin splittings at $-$K and K are 64.93 and 83.61 meV, respectively. We further calculate the Berry curvature when the system is properly hole doped, shifting the Fermi energy to the middle of $-$K and K valleys. For the tensile strain, 4$\%$ for example, the $\Omega_{z}$ reaches its maximum at $-$K point and the difference between $\Omega_{z}$($-$K) and $\Omega_{z}$(K) is enhanced by tensile strain [Fig. S5], compared with Mn$_{2}$BrI monolayer without strain [Fig. 3(c)]. Since the anomalous Hall conductivity is determined by the summation of the Berry curvatures of all the occupied states over the whole Brillouin zone, it is expected that the larger difference between $\Omega_{z}$($-$K) and $\Omega_{z}$(K) would lead to a larger anomalous Hall conductivity.

\subsubsection{Piezomagnetism of Mn$_{2}$BrI monolayer}

\begin{figure}
	\includegraphics[width=0.9\columnwidth]{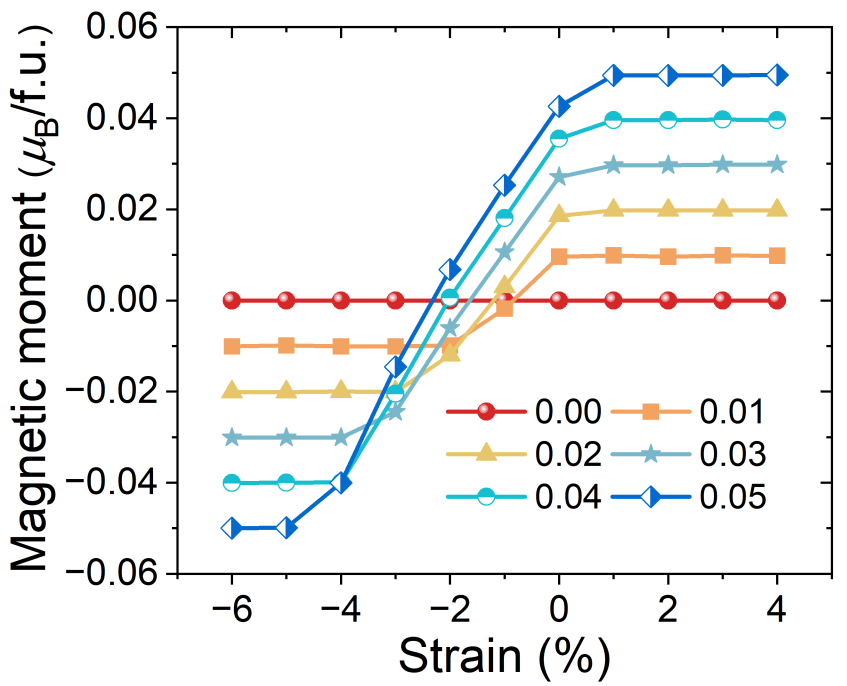}\caption{\label{Fig6} Net magnetic moments (in unit of $\mu_{\mbox{\scriptsize{B}}}$/f.u.) of Mn$_{2}$BrI monolayer as a function of the applied biaxial strain for different hole concentrations (in unit of hole/f.u.).}
\end{figure}

Since the valence band extrema at $-$K, K, and $\Gamma$ points of Mn$_{2}$BrI monolayer can be effectively tuned by external strain and the spin polarization at $\Gamma$ point is different from those at $-$K and K points, we can expect that piezomagnetism could exist in Mn$_{2}$BrI monolayer. Net magnetic moment ($m$) can be obtained by integrating the spin density within the energy range from negative infinity to the Fermi level, which is expressed as $m=\int_{-\infty}^{E_F}(\rho\uparrow-\rho\downarrow)dE$, where $E_{F}$ is the Fermi energy, and $\rho$$\uparrow$ and $\rho$$\downarrow$ represent the density of states (DOS) from spin-up and spin-down channels, respectively. For the pristine Mn$_{2}$BrI monolayer, although the band structure is spin split, the integration of spin-up and spin-down parts of the DOS cancel each other out, therefore the net magnetic moment is zero. However, when the Fermi energy moves downward by hole doping, there will be more spin-down DOS unoccupied owning to the spontaneous spin splitting as well as the valley polarization of the valence bands near the Fermi energy, as demonstrated in Fig. S6(c). Consequently, nonzero net magnetic moment could be obtained in Mn$_{2}$BrI monolayer. For hole doping with concentration of 0.02 hole/f.u., a magnetic moment of 0.02 $\mu_{\mbox{\scriptsize{B}}}$/f.u. could be obtained for the unstrained system. Moreover, the net magnetic moment increases monotonically as a function of the doping concentration we considered, which is shown in Fig. 6.

As mentioned above, for biaxial compressive strain, the band extremum at $\Gamma$ point is elevated gradually when the strain is increased, while the band extrema at $-$K and K points shift downward [Figs. 5(a) and (b)]. As a result, the energy difference among $-$K, $\Gamma$, and K points increases when the compressive strain is increased. For a given doping concentration (0.02 hole/f.u. as an example), the unoccupied valence band extremum changes from spin-down channel dominated to spin-up channel dominated. Consequently, the sign of net magnetic moment reverses from positive to negative when the applied compressive strain is gradually increased. When the compressive strain increases up to about $-3\%$ strain, the spin-down band extrema at both $-$K and K points move below the Fermi level and the unoccupied valence band extremum completely comes from the spin-up channel. Therefore, for hole doping concentration of 0.02 hole/f.u., the net magnetic moment saturates at the $-3\%$ strain, as shown in Fig. 6. When tensile strain is applied, the band extremum at $\Gamma$ point moves downward, while the valley polarization at $-$K and K points is increased. The unoccupied band extrema completely come from the spin-down channel, as long as the hole doping concentration is not so high to move the Fermi level reaching the spin-up band at $-$K point. Therefore the net magnetic moments are positive when the tensile strain is applied.

\begin{figure}
	\includegraphics[width=1.0\columnwidth]{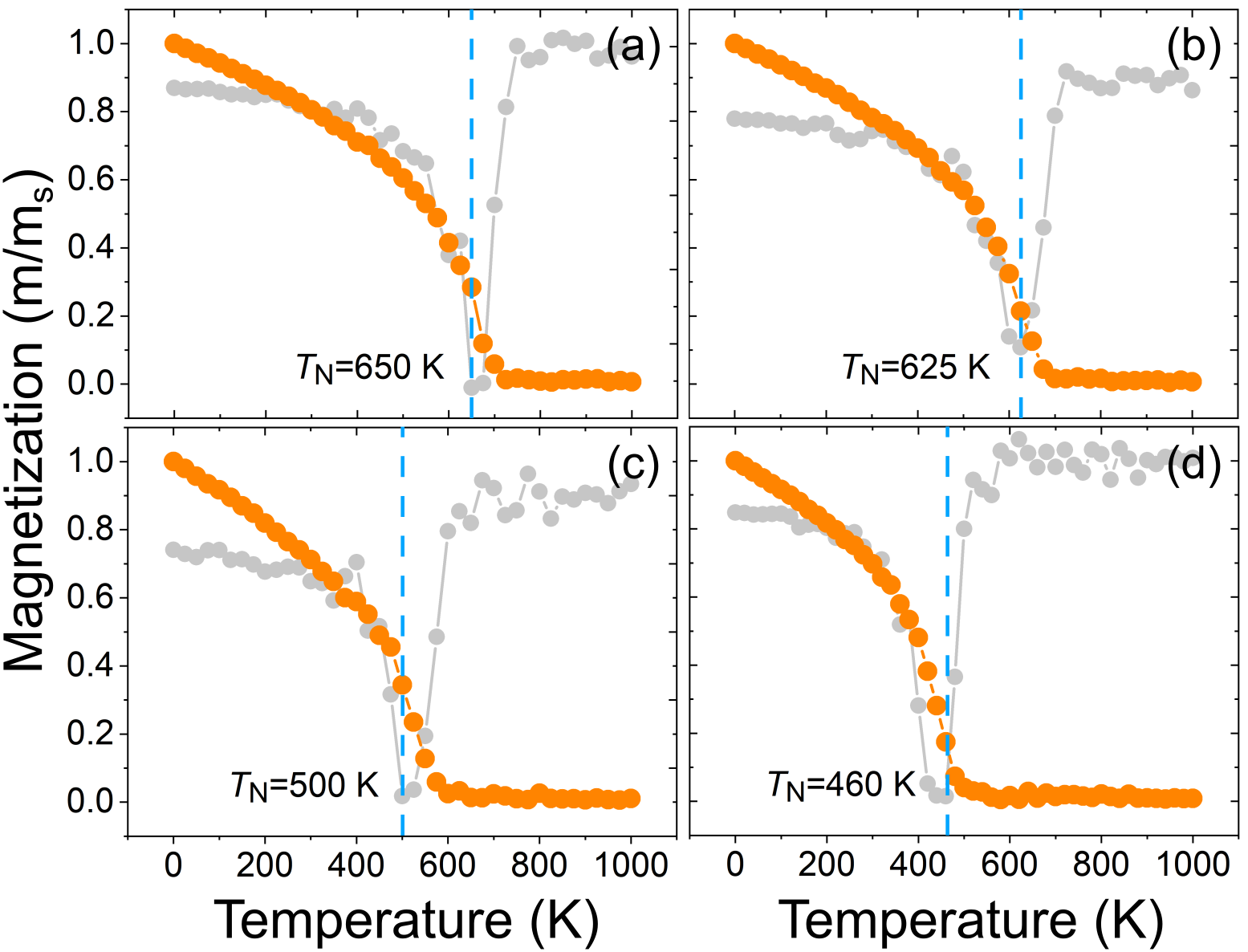}\caption{\label{Fig7} Temperature-dependent sublattice magnetization of (a) $-4\%$ strained and 0.02 hole/f.u. doped, (b) $-2\%$ strained and 0.02 hole/f.u. doped, (c) pristine, and $4\%$ strained and 0.02 hole/f.u. doped Mn$_{2}$BrI monolayer. The light gray lines represent the results of the gradient $dm/dT$, the minimum of which are taken as estimation of N\'{e}el temperature $T_{N}$.}
\end{figure}

For a particular compressive strain, $-4\%$ for example, the net magnetic moment increases monotonically as a function of the doping concentration until the Fermi level moves to the band extremum at $-$K point. When the doping concentration is further increased, the spin-down channel will contribute to the unoccupied bands, leading to the slight decrease of the net magnetic moment. For a particular doping concentration, we can see from Fig. 6 that the net magnetic moment changes from a saturated positive value at tensile strain to a saturated negative value at compressive strain, exhibiting a giant piezomagnetic effect. The magnitude of the net magnetic moments is comparable to those reported in altermagnets V$_{2}$Se$_{2}$O \cite{NatCommun-V2Se2O-2021} and V$_{2}$SeTeO \cite{NanoLett-V2SeTeO-2024} monolayers, for which the piezomagnetism occurs only upon the application of uniaxial or unequal biaxial strains.

\subsection{N\'{e}el temperature of Mn$_{2}$BrI monolayer}

For practical application of the magnetic material, the magnetic transition temperature is an important factor needing to be considered. The N\'{e}el temperature of Mn$_{2}$BrI monolayer was estimated based on the method of Green's functions (Fig. 7). For pristine Mn$_{2}$BrI monolayer, the N\'{e}el temperature is calculated to be 500 K [Fig. 7(c)], higher than the room temperature. Since the MAE is significantly increased by combining the methods of compressive strain and hole doping, the N\'{e}el temperature can be further increased. For example, for system with compressive strain of $-4\%$ and hole doping of 0.02 hole/u.c., the N\'{e}el temperature is enhanced to be 650 K [Fig. 7(a)]. For the tensile strain, however, the N\'{e}el temperature is slightly decreased [Fig. 7(d)], but it remains above room temperature.

\section{Conclusion}

In summary, we predict by first-principles calculations that valley polarization and spontaneous spin splitting can be realized in the Janus Mn$_{2}$BrI monolayer, which is a fully compensated ferrimagnet with in-plane magnetic anisotropy and above-room-temperature N\'{e}el temperature. Spontaneous spin splitting is observed in the absence of SOC, due to the built-in electric field arising from the Janus structure. When SOC is considered, valley polarization emerges for an out-of-plane N\'{e}el vector. When the system is properly hole doped, both the perpendicular MAE and AVH effect can be realized. Furthermore, the MAE of Mn$_{2}$BrI monolayer is significantly enhanced by combining the methods of strain and hole doping. The valence band extrema of Mn$_{2}$BrI monolayer can be effectively tuned by external strain and giant piezomagnetism can be achieved. The fully compensated ferrimagnetism as well as the tunability of valley and magnetic properties of Mn$_{2}$BrI monolayer makes it a promising candidate for future applications in valleytronics and spintronics.

\section{Acknowledgement}

This work was supported by the National Natural Science Foundation of China (Grants Nos. 12474112, 12241405, and 12274411), the National Key R\&D Program of China (Grant No. 2024YFB3614100), the Basic Research Program of the Chinese Academy of Sciences Based on Major Scientific Infrastructures (Grant No. JZHKYPT-2021-08), and the CAS Project for Young Scientists in Basic Research (Grant No. YSBR-084). The calculations were completed on the HPC Platform of Hefei University of Technology.

\end{document}